\documentclass[reprint,amsmath,amssymb,aps,twocolumn,superscriptaddress]{revtex4-1}
\usepackage{graphicx}
\usepackage{dcolumn}
\usepackage{bm}
\usepackage{float}
\usepackage{amssymb}
\usepackage{color}
\usepackage{multirow}
\usepackage{stmaryrd}
\usepackage[colorlinks,linkcolor=blue,urlcolor=blue,citecolor=blue]{hyperref}

\begin{document}


\title{Highly anisotropic transient optical response of charge density wave order in ZrTe$_3$}

\author{Li Yue}
\email{lilyyue@pku.edu.cn}
\affiliation{International Center for Quantum Materials, School of Physics, Peking University, Beijing 100871, China}
\author{Amrit Raj Pokharel}
\affiliation{Institute of Physics, Johannes Gutenberg University, 55128 Mainz, Germany}
\author{Jure Demsar}
\affiliation{Institute of Physics, Johannes Gutenberg University, 55128 Mainz, Germany}
\author{Sijie Zhang}
\affiliation{International Center for Quantum Materials, School of Physics, Peking University, Beijing 100871, China}
\author{Yuan Li}
\affiliation{International Center for Quantum Materials, School of Physics, Peking University, Beijing 100871, China}

\author{Tao Dong}
\affiliation{International Center for Quantum Materials, School of Physics, Peking University, Beijing 100871, China}
\author{Nanlin Wang}
\email{nlwang@pku.edu.cn}
\affiliation{International Center for Quantum Materials, School of Physics, Peking University, Beijing 100871, China}
\affiliation{Beijing Academy of Quantum Information Sciences, Beijing 100913, China}


\begin{abstract}

Low dimensionality in CDW systems leads to anisotropic optical properties, in both equilibrium and non-equilibrium conditions. Here we perform polarized two-color pump probe measurements on a quasi-1D material ZrTe$_3$, in order to study the anisotropic transient optical response in the CDW state. Profound in-plane anisotropy is observed with respect to polarization of probe photons. Below $T_\mathrm{CDW}$ both the quasi-particle relaxation signal and amplitude mode (AM) oscillation signal are much larger with $\mathbf{E}_\mathrm{pr}$ nearly parallel to $a$ axis ($\mathbf{E}_\mathrm{pr} \parallel a$) than for $\mathbf{E}_\mathrm{pr}$ parallel to $b$ axis ($\mathbf{E}_\mathrm{pr} \parallel b$). This reveals that $\mathbf{E}_\mathrm{pr} \parallel a$ signal is much more sensitive to the variation of the CDW gap. Interestingly, the lifetime of the AM oscillations observed with $\mathbf{E}_\mathrm{pr} \parallel b$ is longer than $\mathbf{E}_\mathrm{pr} \parallel a$. Moreover, at high pump fluence where the electronic order melts and the AM oscillations vanish for $\mathbf{E}_\mathrm{pr} \parallel a$ , the AM oscillatory response still persists for $\mathbf{E}_\mathrm{pr} \parallel b$. We discuss possible origins that lead to such unusual discrepancy between the two polarizations.

\end{abstract}

\pacs{Valid PACS appear here}
\maketitle



Low dimensionality plays a key role for the formation of charge density waves (CDWs). CDWs usually appear in quasi-1D or -2D materials \cite{MonceauAdvPhys2012,RossnagelJPhysCondMat2011}, where the low-dimensional electronic structure supports good Fermi surface nesting. The electronic susceptibility, especially for quasi-1D systems, tend to diverge at the nesting wave vector 2$\mathbf{k}_F$ \cite{gruner2018density,gruner1988dynamics}, resulting in electronic instability and periodic modulation of the conducting electrons. Another important factor for CDW formation is electron-phonon coupling. At 2$\mathbf{k}_F$, phonon modes tend to soften, leading to a new periodic modulation of the lattice structure \cite{gruner1988dynamics,gruner2018density,PougetPRB1991,HoeschPRL2009}.

A typical experimental characterization of CDW phase is the measurement of optical properties, in both equilibrium and non-equilibrium conditions. Optical conductivity spectroscopy is a powerful tool to identify the single-particle gap in the CDW state, where profound spectral weight redistribution happens in the low-energy range \cite{LiPRB2022,LinPRB2020,ChenPRB2014,ChenPRL2017}. CDWs also induce new collective excitations consisting of amplitude (AM) and phase modes. The AM excitations involves ionic displacement around the periodically distorted atomic displacements, and can be observed in Raman spectroscopy as a soft optical mode with $A_\mathrm{1g}$ symmetry \cite{HuPRB2015,SnowPRL2003}. Besides these time-averaging spectroscopies which detect the equilibrium properties, ultrafast pump probe techniques have been widely utilized to study the non-equilibrium dynamical evolution of the system after optical excitation \cite{LiPRB2022,LinPRB2020,ChenPRB2014,ChenPRL2017,DemsarPRL1999,DemsarPRB2002,SchaeferPRB2014}. The single-particle gap opening and the appearance of collective AM are manifested in the time-domain spectra as changes in quasi-particle (QP) relaxation signals and coherent oscillations, respectively. Moreover, time-resolved probes can provide much more information of the system that is inaccessible in the equilibrium state, including non-thermal ultrafast phase transitions driven by pump excitation \cite{TomeljakPRL2009,SchmittScience2008}, and selective detection of the order parameter components during the relaxation process\cite{TomeljakPRL2009,SchaferPRL2010}.

The low dimensionality of CDW systems can lead to strongly anisotropic optical responses. Optical conductivity experiments have built a thorough understanding of the anisotropy in the equilibrium state. For in-plane optical conductivities of many quasi-1D CDW systems, the gap signature mainly appears with the polarization direction along the in-plane projection of 2$\mathbf{k}_F$ vector (which is often the structural chain direction) \cite{LiPRB2022,PerucchiPRB2004}. In contrast, for the non-equilibrium dynamical relaxation process, the anisotropy of the transient optical response was seldom studied. Such studies would be meaningful for clarifying the dynamical evolution of CDW state, and expand the understanding of the cooperative phenomena. Here we report polarized femtosecond pump probe measurements on a prototypical quasi-1D CDW material ZrTe$_3$, which has not been previously studied with pump probe methods. Our data reveal interesting and pronounced anisotropic transient optical responses of the system.

ZrTe$_3$ is a quasi-1D CDW metal with $T_\mathrm{CDW}$=63 K.
It belongs to the space group $P2_1/m$, with $a=5.89$ \r{A}, $b=3.93$ \r{A}, $c=10.09$ \r{A}, $\beta=97.8^{\circ}$, $\alpha=\gamma=90^{\circ}$ \cite{HuPRB2015,StoweJSolidStateChem1998}. The structure consists of prisms stacking along $b$ axis. The sample has stripes-like $ab$ surface, with the long edge parallel to $b$ axis and short edge parallel to $a$ axis. The ordering wave vector $\mathbf{q}_\mathrm{CDW}=(0.07 \mathbf{a}^*,0,0.33 \mathbf{c}^*)$ corresponds to the nesting of the quasi-1D Fermi sheets near the Brillouin zone boundary \cite{YokoyaPRB2005,HuPRB2015,YueNatCommu2020}. Unlike other quasi-1D CDW materials such as NbSe$_3$, K$_{0.3}$MoO$_3$, etc., $\mathbf{q}_\mathrm{CDW}$ is perpendicular to the structural chain direction. According to ARPES experiments, the quasi-1D Fermi sheets are partially gapped around the Brillouin zone $D$ point\cite{YokoyaPRB2005,HoeschPRL2019}, where the electron-phonon interaction is strongest \cite{HuPRB2015}. Optical conductivity measurements showed the anisotropic gap feature which is more prominent along the in-plane $a$ direction \cite{PerucchiEurPhysJB2005}, with a depletion of spectral weights below 1000 cm$^{-1}$.

\begin{figure}[htbp]
	\begin{center}
		\includegraphics[clip, width=0.5\textwidth]{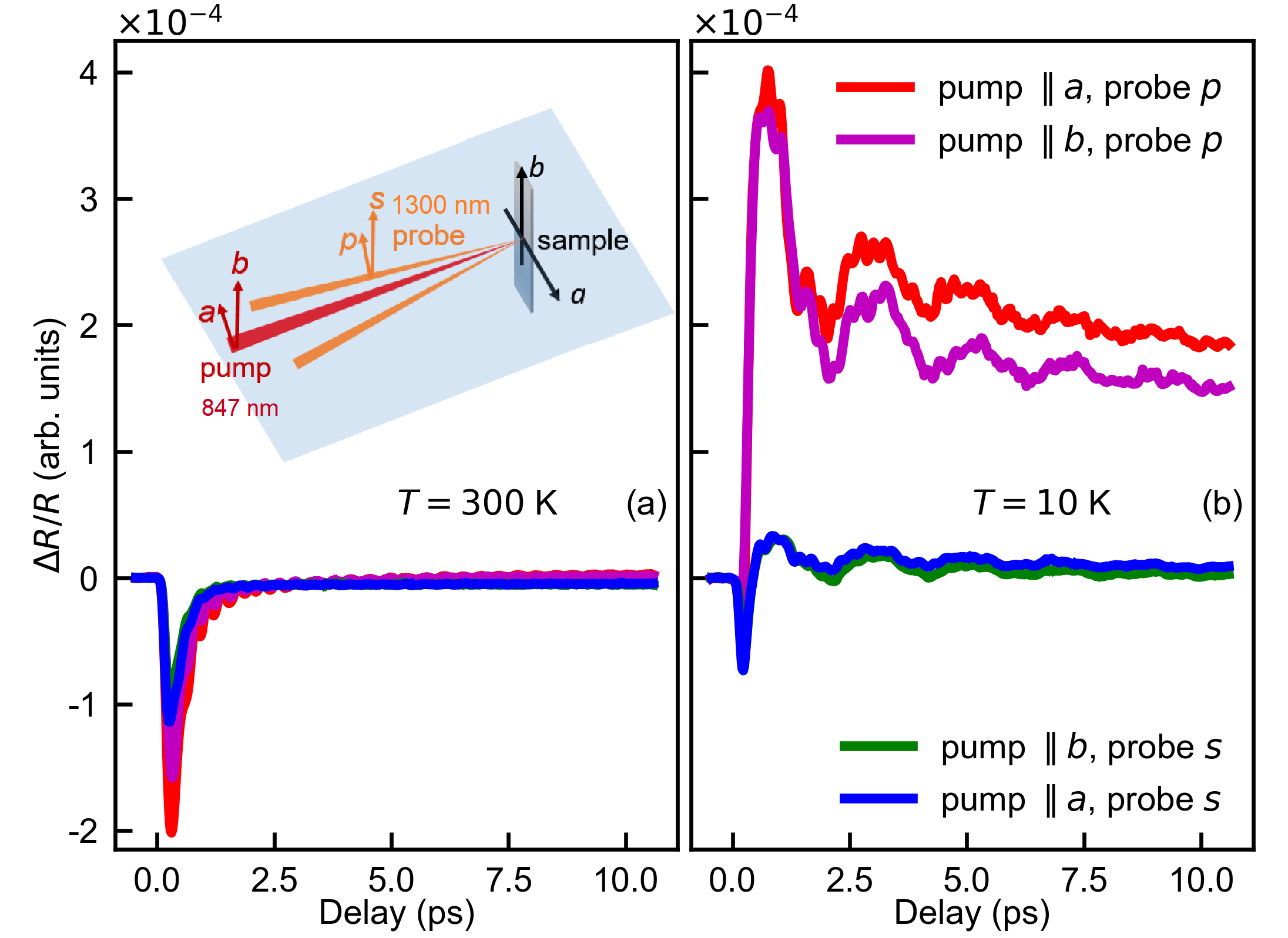}\\[1pt] 
		\caption{(a) and (b) Time resolved reflectivity  under different polarization conditions at room temperature and 10 K, respectively, with pump fluence $\sim 60$ $\mu$J/cm$^2$. The inset of panel depicts the reflection geometry of the experiment. The $a$ axis lies in the horizontal plane of incidence, and the $b$ axis is along the vertical direction.}
		\label{fig:1}
	\end{center}
\end{figure}

We performed two-color pump probe reflection measurements on the $ab$ surface of ZrTe$_3$ single crystal \cite{SM}. Fig.~\ref{fig:1} displays the polarization-dependent transient reflectivity changes at room temperature and at 10 K. It is clear that the polarization of probe photons makes the major difference to the signal amplitudes. In the CDW state (Fig.~\ref{fig:1}(b)), the difference between the two probe polarizations is much more profound than in the normal state (Fig.~\ref{fig:1}(a)). Since the polarization dependent spectra are dominated by the $\mathbf{E}_\mathrm{pr}$ direction, we focus on the results obtained with $\mathbf{E}_\mathrm{pu} \parallel a$. Considering the rather small incident angle of the probe beam, the $s$ and $p$ polarizations can be approximated as nearly parallel to $b$ ($\mathbf{E}_\mathrm{pr} \parallel b$) and $a$ ($\mathbf{E}_\mathrm{pr} \parallel a$) axis.

\begin{figure}[htbp]
	\begin{center}
		\includegraphics[clip, width=0.5\textwidth]{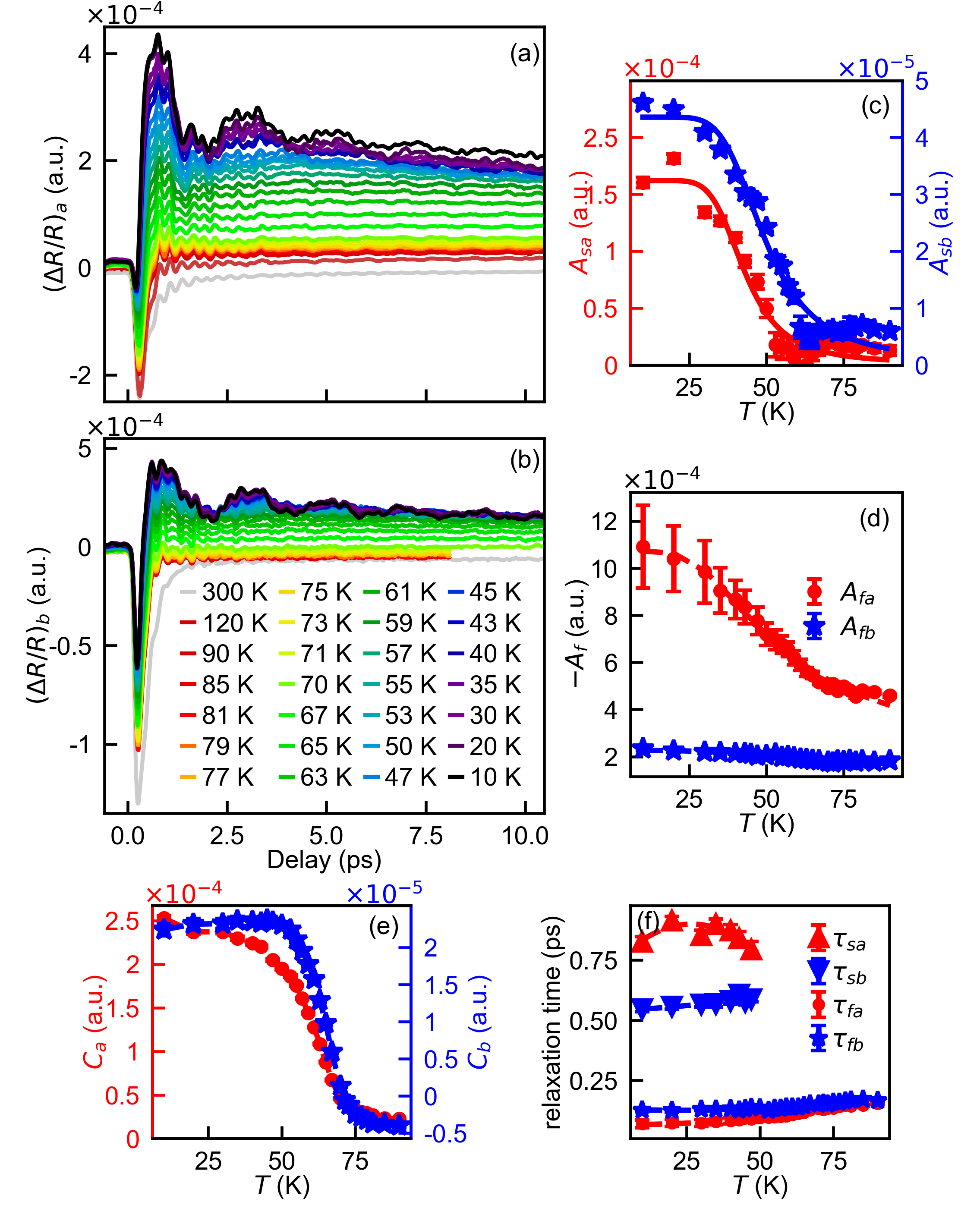}\\[1pt] 
		\caption{(a) and (b) $\Delta R/R$ as a function of time delay at different temperatures with pump fluence $\sim 60$ $\mu$J/cm$^2$. (c-f) Temperature dependence of the several parameters characterizing the QP relaxation dynamics. The red and blue markers correspond to spectra obtained with $\mathbf{E}_\mathrm{pr} \parallel a$ and $\mathbf{E}_\mathrm{pr} \parallel b$, respectively. The solid lines in (c) are the fitted curves with RT model using $T$-independent gap \cite{DemsarPRL1999YBCO}. The dashed lines are guides-to-the-eye.}
		\label{fig:2}
	\end{center}
\end{figure}

Figures~\ref{fig:2}(a-b) present the temperature evolution of the transient reflectivity change with $\mathbf{E}_\mathrm{pr}\parallel a$ and $\mathbf{E}_\mathrm{pr}\parallel b$, respectively. In Fig.~\ref{fig:2}(a), at 300 K the relaxation dynamics is given by a fast decaying negative signal with a small long-lived background signal extending far over the 10 ps time-window. As temperature is decreased towards $T_\mathrm{CDW}$, the long-lived background signal changes sign. Across the CDW transition, a second exponential decay process with positive sign and longer relaxation time becomes apparent while the amplitude of the long-lived background response rapidly increases with decreasing temperature. Upon cooling to the lowest temperature, the two exponentially decaying signals, as well as the long-lived signal are further enhanced, and the oscillatory response due to the CDW amplitude mode with $\sim 2$ ps period becomes apparent. The spectra for $\mathbf{E}_\mathrm{pr} \parallel b$ (Fig.~\ref{fig:2}(b)) show a similar trend, but with substantially lower signal amplitudes.

The QP relaxation dynamics can be fit by a convolution of a Gaussian response function and the following model function
\begin{align}\label{Eq:1}
(\frac{\Delta R}{R})_i=A_{fi} \mathrm{e}^{-\frac{t}{\tau_{fi}}}+A_{si} \mathrm{e}^{-\frac{t}{\tau_{si}}}+B_{i}t+C_i
\end{align}
\noindent  Here $i=a$ or $b$ denotes the $\mathbf{E}_\mathrm{pr}$ polarization, $A_{fi}$ and $\tau_{fi}$ are the amplitude and relaxation time of the fast negative exponential decay seen at all temperatures, $A_{si}$ and $\tau_{si}$ are the the amplitude and relaxation time of the picosecond positive component arising in the CDW phase. The long-lived background signal should, in principle, be fitted with a third exponential decay with very long relaxation time. However, since we are focusing on the short time dynamics on the timescales of about 10 ps, we approximate the long lived response by $B_{i}t+C_i$.

In the pump probe experiment, the pump pulse injects QPs into the system. If a gap exists near the Fermi level, photo-generated carriers will accumulate above the gap, the presence of the latter resulting in a relaxation bottleneck. However, in ZrTe$_3$, only parts of the Fermi surface are gapped by the CDW formation \cite{YokoyaPRB2005}. Therefore, part of the photo-generated carriers may rapidly relax to ungapped region of the Fermi surface and relax via normal electron-phonon thermalization process. The two exponentially decaying components in Eq.~\ref{Eq:1}, with the faster one being present at all temperatures,  likely stem from the two relaxation channels.

The amplitudes of both exponentially decaying components exhibit anomalies near $T_\mathrm{CDW}$. In Fig.~\ref{fig:2}(c), $A_{si}$ is nearly zero at high temperatures, and increases rapidly below $T_\mathrm{CDW}$. $A_{fi}$ are also enhanced in the CDW state (Figs.~\ref{fig:2}(d)), however they remain pronounced far above $T_\mathrm{CDW}$. Compared to the value at 10 K $A_{fi}$ above $T_\mathrm{CDW}$ is about $40 \%$ of its low temperature value for $\mathbf E_\mathrm{pr} \parallel a$ and $80 \%$ for $\mathbf E_\mathrm{pr} \parallel b$, consistent with the assignment of the fast decay component to the normal relaxation processes in the un-gapped regions. The relaxation times $\tau_{fi}$ are on the order of $\sim$ 100 fs, displaying no anomalies near $T_\mathrm{CDW}$, consistent with this scenario. The long-lived signal $C_i$ also changes dramatically near $T_\mathrm{CDW}$. This signal may be in part related to the bolometric response as well as to the localized in-gap carriers \cite{DemsarPRL1999,DemsarPRB2002}, whose contribution  is reduced above $T_\mathrm{CDW}$, consistent with the pronounced drop in $C_i$ at temperatures above $T_\mathrm{CDW}$.

Previously, the anisotropic optical response to the CDW gap formation was mainly studied by optical conductivity measurement of the equilibrium state. Here, in quasi-1D systems, the gap feature is more prominent along the Fermi nesting direction and is weak along the perpendicular direction. In the near-infrared range, far above the CDW gap, the optical conductivity is very weakly influenced by the CDW transition and is usually not resolved in the equilibrium spectra. In the optical pump probe experiment, however, thanks to the high sensitivity of the technique, one can measure the pump induced reflectivity changes due to the photoinduced gap suppression, even if they are much smaller than $10^{-4}$. Our polarized pump-probe study, even at 0.95 eV (1300 nm) probe photon energy, clearly demonstrates a profound anisotropic transient optical response in the CDW state, where $\mathbf{E}_\mathrm{pr}\parallel a$ signal in the CDW state is about an order of magnitude larger than for $\mathbf{E}_\mathrm{pr}\parallel b$. At temperatures above $T_\mathrm{CDW}$, however, the difference between the responses in the two probe polarizations is quite small (see Fig. 1(a) and Fig. 2(d)). We note that such an anisotropy of the response in the CDW state should not be taken for granted in quasi-1D CDW systems. E.g., in studies of K$_{0.3}$MoO$_3$, the anisotropy between the probe polarizations along the chain or perpendicular to the chain directions is quite weak (see Supplementary material in Ref. \cite{SchaferPRL2010}).

\begin{figure}[htbp]
	\begin{center}
		\includegraphics[clip, width=0.5\textwidth]{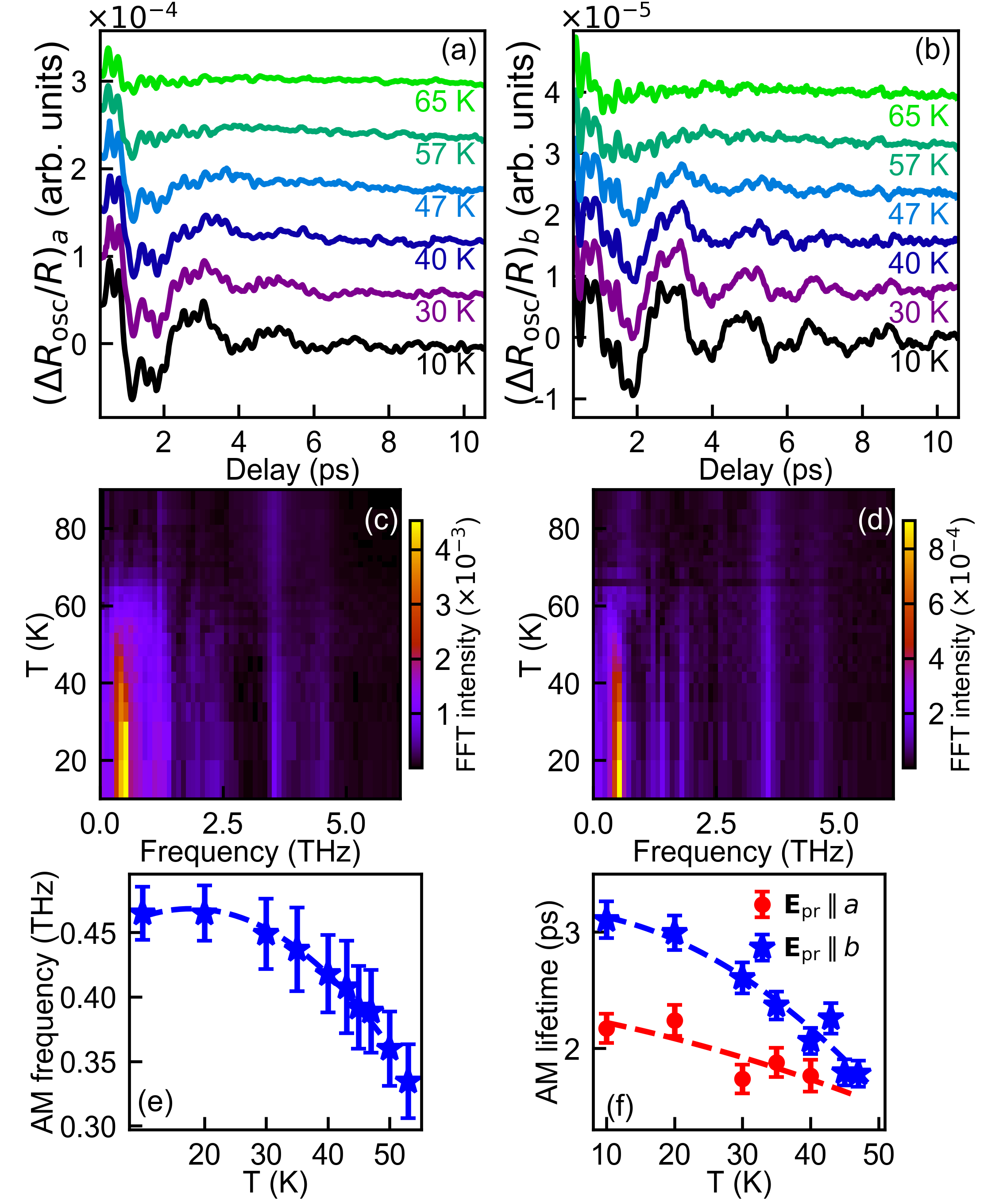}\\[1pt] 
		\caption{(a) and (b) The oscillation part of transient reflectivity change for $\mathbf{E}_\mathrm{pr} \parallel a$ and $\mathbf{E}_\mathrm{pr} \parallel b$ with pump fluence. Lines are vertically shifted for clarity. (c) and (d) correspond to the Fast Fourier transformation of the oscillation part for $\mathbf{E}_\mathrm{pr} \parallel a$ and $\mathbf{E}_\mathrm{pr} \parallel b$, respectively. (e) The temperature dependence of AM peak frequency obtained by fitting the peak position in the FFT spectra of panel (d). (f) The lifetime of AM oscillation for different probe photon polarizations, obtained by fitting the time-domain oscillation.}
		\label{fig:3}
	\end{center}
\end{figure}

Figures \ref{fig:3}(a-b) display the oscillatory component of the spectra along two polarizations, obtained by subtracting the fitted QP relaxation curve of Eq.~\ref{Eq:1} from the raw traces in Figs. \ref{fig:2}(a-b). Figs. \ref{fig:3}(c-d) present the corresponding fast Fourier transformation. Several distinct modes are observed at around 0.47, 1.74, 3.5, 4.5, 6.55 THz, among which the 0.47 THz mode, disappearing above $T_\mathrm{CDW}$, shows clear signatures of the amplitude mode. As the temperature rises, the AM intensity decays and the mode softens (Fig. \ref{fig:3}(e)), consistent with the Raman measurements \cite{HuPRB2015}. Comparing the two polarizations, the AM oscillation intensity for $\mathbf{E}_\mathrm{pr}\parallel a$ is much stronger than that for $\mathbf{E}_\mathrm{pr}\parallel b$. Since the AM collective dynamics involves the oscillation of the CDW gap \cite{SchmittScience2008,SchmittNewJPhys2011}, the larger AM amplitude for $\mathbf{E}_\mathrm{pr}\parallel a$ implies that $\mathbf{E}_\mathrm{pr}\parallel a$ provides a much larger response to the photoinduced variation of the CDW gap. This is consistent with the larger QP relaxation signal for the probe polarized along the $a$-axis due to the gap formation below $T_\mathrm{CDW}$, as discussed in the preceding section.

The oscillatory AM signal can be described with a damped oscillator
\begin{align}\label{Eq:2}
\frac{\Delta R_\mathrm{osc}}{R}(t)=A_\mathrm{AM}\mathrm{e}^{-t/\tau_\mathrm{AM}}\mathrm{sin}(\omega_\mathrm{AM}t+\phi)
\end{align}
where $A_\mathrm{AM}$, $\tau_\mathrm{AM}$, $\omega_\mathrm{AM}$, $\phi$ correspond to the amplitude, lifetime, frequency and phase. An interesting result of our measurements is that the AM lifetime also exhibits anisotropy (see Fig. \ref{fig:3}(f)): the lifetime is shorter for $\mathbf{E}_\mathrm{pr}\parallel a$ than for $\mathbf{E}_\mathrm{pr}\parallel b$. This difference can also be directly seen in the time-domain traces, where the oscillations in Fig. \ref{fig:3}(b) extend longer than in Fig. \ref{fig:3}(a). We discuss the possible origins of this observation later.

\begin{figure}[htbp]
	\begin{center}
		\includegraphics[clip, width=0.5\textwidth]{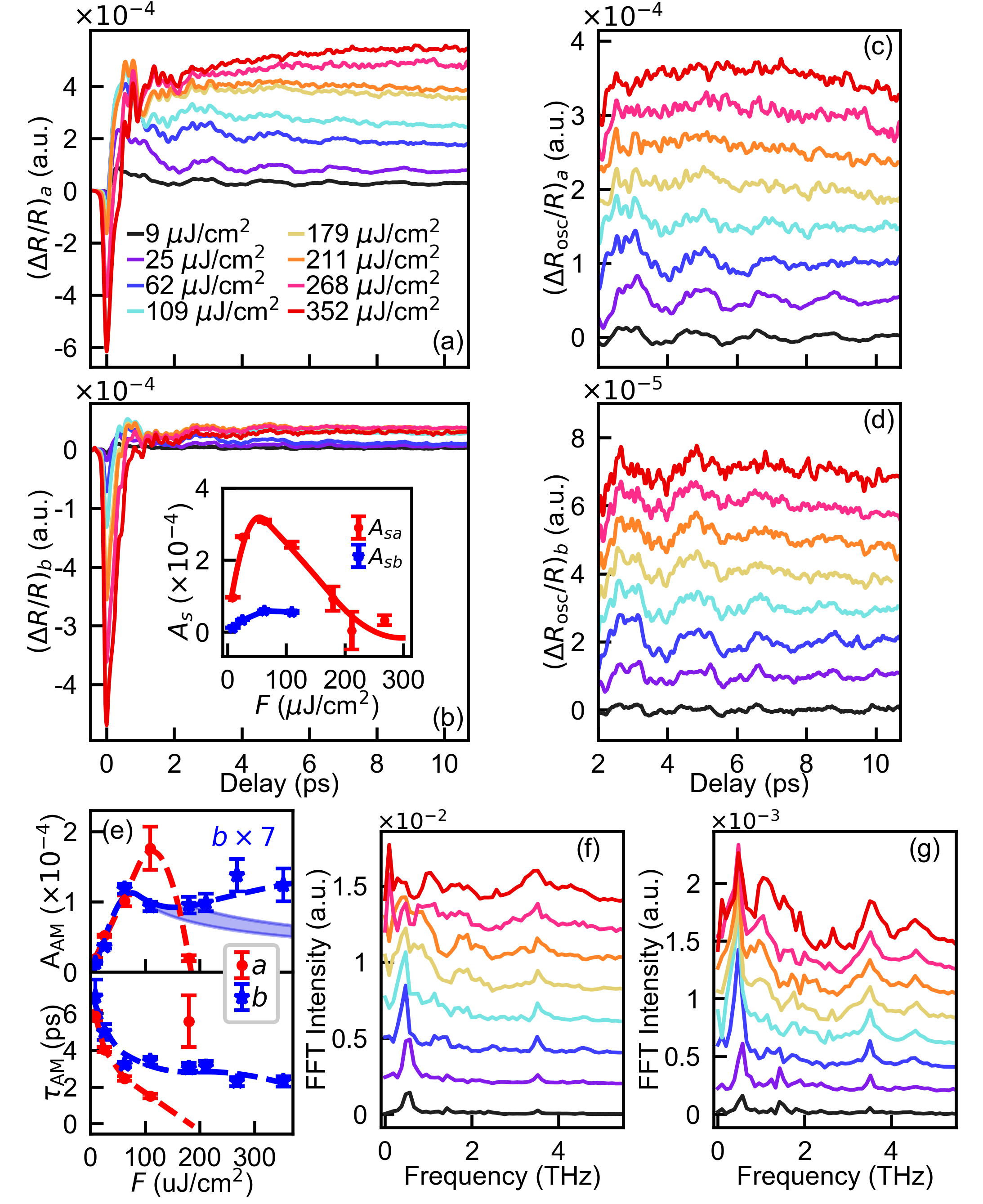}\\[1pt] 
		\caption{(a) and (b) The transient reflectivity change at different pump fluence at 10 K, with $\mathbf{E}_\mathrm{pr} \parallel a$ and $\mathbf{E}_\mathrm{pr} \parallel b$ respectively. The inset of (b) shows the amplitude of the slower positive exponential at different fluence. For $\mathbf{E}_\mathrm{pr} \parallel b$ the we can only obtain reasonable fitting parameter $A_{sb}$ below 179 $\mu$J/cm$^2$. (c) and (d) The oscillation part of the time-domain spectra of (a) and (b), data is taken from 2 ps to avoid the residual quasi-particle relaxation signal which is not perfectly subtracted. (e) The fluence dependence of the amplitude (upper panel) and lifetime (lower panel) of the AM oscillation, obtained by fitting the time-domain oscillation spectra with Eq. \ref{Eq:2}. The shaded region in the upper panel expected intensity range in the condition that the existence of AM signal at high fluence with $\mathbf{E}_\mathrm{pr} \parallel b$ is purely induced by the mismatch of penetration depth \cite{SM}. (f) and (g) The Fast Fourier transformation of the time-domain oscillation signal for different probe polarizations.}
		\label{fig:4}
	\end{center}
\end{figure}

We also studied the pump fluence dependence of the dynamics at 10 K (Figs. \ref{fig:4}(a-b)). By increasing the pump fluence, the system will go from a linear, weak perturbation, regime to quenching the CDW state. This has been observed in other CDW materials, showing that strong perturbation leads to an ultrafast melting of the electronic order where the gap is closed and the amplitude mode amplitudes saturate and eventually vanish \cite{TomeljakPRL2009,SchmittScience2008}. In ZrTe$_3$, with increasing fluence, the fast negative exponential increases monotonically, but becomes less anisotropic, as expected from the T-dependence measurements. The slower positive exponentially decaying component, which corresponds to the QP relaxation across the CDW gap, also grows linearly at first but then fades away (Figs. \ref{fig:4}(b) inset). Such changes of QP relaxation signals are consistent with ultrafast melting of the electronic order. At high pump fluence, a broad hump appears on the higher-frequency side of the AM peak in the FFT spectra, which may be related to enhanced anharmonicity under strong perturbation. Along with the melting of electronic order, the AM oscillation gradually vanishes for $\mathbf{E}_\mathrm{pr}\parallel a$ (Figs. \ref{fig:4}(c, f)). Interestingly, for $\mathbf{E}_\mathrm{pr}\parallel b$ the AM oscillation exist till the highest fluences (Figs. \ref{fig:4}(d, g)).

Generally, one would expect the amplitude mode to saturate and subsequently disappear as the electronic order is melted by high fluence, regardless of the probe photon polarization. Yet here, the AM oscillation for $\mathbf{E}_\mathrm{pr}\parallel b$ survive above $\approx$ 100 $\mu$J/cm$^2$, with its amplitude and lifetime nearly independent on fluence above $\sim$ 100 $\mu$J/cm$^2$ (Fig. \ref{fig:4}(e)). The different fluence dependence of the AM amplitudes between two probe polarizations, as well as the different lifetimes shown in Fig. \ref{fig:3}(f), have not been reported thus far. In the following we discuss the possible origins that may be responsible for these observations.

First, given the non-linearity of the AM response, we consider the influence of penetration depths mismatch. The optical penetration depths are determined from the reported reflectivity data by Herr \textit{et al.} \cite{HerrSyntheticMetals1986}, using Kramers-Kronig analysis. The penetration depth at 1300 nm probe wavelength is 28.2 nm for $\mathbf{E}_\mathrm{pr}\parallel a$ and 89.2 nm for $\mathbf{E}_\mathrm{pr}\parallel b$ while at 847 nm pump wavelength it is $22.5$ nm for $\mathbf{E}_\mathrm{pu}\parallel a$ and 47.8 nm for $\mathbf{E}_\mathrm{pu}\parallel b$.

The large mismatch can thus be responsible for different fluence dependences of the AM signal. Let us first consider the fluence dependence of the response for the case, where pump and probe mismatch is negligible, or when optical penetration depth is larger than that of the probe. For weak perturbations, the amplitude of the AM signal is usually linearly proportional with the pump fluence. Above the absorbed energy density corresponding to quenching the electronic order, however, the signal will saturate \cite{TomeljakPRL2009} and eventually disappear.
Indeed, this scenario well accounts for our observations with $\mathbf{E}_\mathrm{pr}\parallel a$. For the case, where the pump penetration depth is shorter that the probe penetration depth, i.e., for  $\mathbf{E}_\mathrm{pr}\parallel b$, however, the situation is more complicated. Beyond the linear response regime, the high incident pump fluence result in quenching of the electronic order in the surface layer given by the pump penetration depth. However, for the deeper regions (still probed by the probe), the absorbed energy density is still below the CDW melting threshold, resulting in a contribution to the reflectivity transient of the perturbed CDW state. Indeed, our observations with $\mathbf{E}_\mathrm{pr}\parallel b$ are qualitatively consistent with this scenario. E.g., the quenched surface region may be responsible for the broad hump feature and broadened phonon peaks in the FFT spectra, with the deeper, weakly perturbed regions giving rise to the AM response far beyond the threshold seen in $\mathbf{E}_\mathrm{pr}\parallel b$ experiment.

In order to check this scenario, we performed quantitative analysis with the above model, presented in the Supplementary material \cite{SM}. The model, taking into account the optical penetration length mismatch, is qualitatively consistent with our experimental observations. However, after quenching of the electronic order in the surface layer, by further increasing the pump fluence, the volume fraction of the quenched CDW state will increase, leading to decrease of the AM signal intensity. According to the model the AM signal intensity should drop faster with increasing fluence as compared to our experimental results.
The predicted dependence of the amplitude of the AM is given by the dashed blue region in Fig. \ref{fig:4} (e) \cite{SM}, showing that for fluences beyond $\approx$ 100 $\mu$J/cm$^2$, the AM signal intensity for $\mathbf{E}_\mathrm{pr}\parallel b$ should start to decay. However, the measured AM signal intensity for $\mathbf{E}_\mathrm{pr}\parallel b$ is nearly constant and may even be increasing with increasing pump fluence. Moreover, we note that the lifetimes of AM oscillations in Fig. \ref{fig:1}(b) do not change much when switching pump polarization to from $a$ to $b$ direction (Supplementary Fig. 2) \cite{SM}, even though the difference in the pump penetration depths is about a factor of two.
Thus, the penetration depth mismatch may not be solely responsible for our observations. We should note, however, that the non-linearity of the AM signal can be more complicated than our simple phenomenological model.

Next we discuss an alternative scenario that may be responsible for the observe anisotropy in the fluence dependent AM response. We first note that the CDW order parameter comprises electronic and lattice parts, and the amplitude mode involves oscillation of the electronic gap \cite{SchmittScience2008,SchmittNewJPhys2011,DolgirevPRB2020} and the periodic lattice displacements \cite{MoorePRB2016}. Usually, the two parts are considered to be adiabatically coupled due to electron-phonon coupling. However, at least on the timescales shorter that the AM period, this assumption may be invalid \cite{SchaferPRL2010,SchaeferPRB2014}. First, in earlier studies, different techniques have been utilized to probe the ultrafast dynamics of the two components of the CDW order. In TbTe$_3$, time-resolved photoemission demonstrated the oscillation of CDW gap at the $\sim 2.4$ THz AM frequency \cite{SchmittScience2008,SchmittNewJPhys2011}. Here, above $\sim 1$ mJ/cm$^2$, the gap is closed by strong excitation and the AM oscillation are no longer resolved. However, the time-resolved X-ray diffraction study on TbTe$_3$ showed that with 3 mJ/cm$^2$ fluence the superlattice peak intensity still oscillates at the AM frequency \cite{MoorePRB2016}. In K$_{0.3}$MoO$_3$, X-ray studies suggest a critical fluence of ~ 1 mJ/cm$^2$ for the collapse of the superlattice \cite{HuberPRL2014}, but optical reflectivity probe showed a critical fluence of $\sim 0.3$ mJ/cm$^2$, where damping and disappearance of the AM oscillation is observed \cite{TomeljakPRL2009}. Finally, at fluences much higher than those needed to collapse the electronic order, a collapse and revival of the periodic lattice modulation has been observed \cite{HuberPRL2014,Neugebauer2019}. Thus, it can be inferred that the non-equilibrium dynamics of the electronic and lattice subsystems are not always strictly entangled. Under strong perturbation, the electronic order is melted and CDW band oscillation vanish (or become to small to detect), but the lattice part may still display well resolved oscillatory dynamics. Such a disentangled dynamics may be responsible for the observed anisotropy in the fluence dependence of AM in ZrTe$_3$, as discussed below.

To describe this scenario, we first point out the fact, that transient reflectivity in the infrared range far above the CDW gap is neither a direct probe of the lattice motion or of the CDW gap. In fact, the reflectivity probes both components with different weights. The weights may depend on the photon energy. E.g., when the probe photon energy is close to the CDW gap, one can expect a dominant contribution to the signal from the CDW gap oscillation, while in the case of a resonance with an interband transition, the probe may be more sensitive to lattice part of the order parameter, where the whole electronic band structure will experience shifts due to the driven atomic motion \cite{SchaeferPRB2014}. The weights may also depend on photon polarizations, given the anisotropic CDW gap which affects the optical conductivities mainly along $a$ direction \cite{PerucchiEurPhysJB2005}. With the highly anisotropic electronic structure, manifested by a large anisotropy in the dielectric function up to the visible range \cite{HerrSyntheticMetals1986}, it is possible that transient reflectivities in the $a$ and $b$ polarizations are mainly probing the collective dynamics of the electronic and lattice subsystems, respectively. Such a scenario is also consistent with our observation of AM at high fluences for $\mathbf{E}_\mathrm{pr}\parallel b$. Here, at high fluence, the disappearance of AM signal for $\mathbf{E}_\mathrm{pr}\parallel a$ corresponds to the strong suppression of the AM oscillations of the electronic gap, while the AM signal for $\mathbf{E}_\mathrm{pr}\parallel b$ corresponds to the dynamics of the lattice part of the order parameter. This scenario may be realized in ZrTe$_3$ given the fact that $\mathbf{E}_\mathrm{pr}\parallel a$ provides much larger response to the variation of the CDW gap, as shown by the profound intensity of the quasiparticle relaxation and the AM oscillation signals in this polarization. For $\mathbf{E}_\mathrm{pr}\parallel a$, the measured AM oscillation is dominated by the dynamical evolution of CDW gap. For $\mathbf{E}_\mathrm{pr}\parallel b$, the signal is not as sensitive to the variation of gap, so the response would be dominated by the AM vibration.
Indeed, such a scenario agrees with the different amplitudes of individual components to transient reflectivity in the two probe polarizations - see Fig. \ref{fig:1}(b). Here for $\mathbf{E}_\mathrm{pr}\parallel a$ the QP exponential decay is very prominent, but for $\mathbf{E}_\mathrm{pr}\parallel b$ the signal is dominated by the coherent AM oscillation.

In summary, we use polarized femtosecond pump probe experiments to investigate the non-equilibrium dynamical optical response in ZrTe$_3$. Profound anisotropy is observed with respect to polarization of probe photons. With $\mathbf{E}_\mathrm{pr}\parallel a$, the spectra present much larger response to the formation of CDW gap below $T_\mathrm{CDW}$ as well as to the variation of CDW gap during AM oscillations. Interestingly, we observe that the AM oscillation signal present different damping lifetime between $\mathbf{E}_\mathrm{pr}\parallel a$ and $\mathbf{E}_\mathrm{pr}\parallel b$. Moreover, the AM oscillation signal for $\mathbf{E}_\mathrm{pr}\parallel b$ is seen to persist to much higher pump fluence than for $\mathbf{E}_\mathrm{pr}\parallel a$. While the results may be attributed to the anisotropy of the pump and probe penetration depths, they may also be linked to the disentangled collective dynamics of electronic and lattice parts of the order parameter. Although numerous optical pump probe measurements have been performed on low-dimensional CDW materials, our results suggest that further, spectrally resolved studies may further our understanding of the cooperative phenomena in CDW materials as well as correlated materials in general.

This work was supported by NSFC (No. 11888101 and NO. 2022YFA1403900). L.Y. further acknowledges support from China Postdoctoral Science Foundation (Grant No. 2021M700257 and Grant No. BX20200016). A.R.P. and J.D. acknowledge support by the Deutsche Forschungsgemeinschaft (DFG, German Research Foundation) - TRR 288 - 422213477 (project B08).

\bibliography{Ref_ZrTe3_pumpprobe}

\clearpage

\end{document}